**Interpersonalized Affordances: A Uses and Gratifications Analysis of AI Companionship**

Dayeon Eom[1*], Julianne Renner[1]
[1]Department of Life Sciences Communication, University of Wisconsin-Madison, 1545 Observatory Drive, Madison, WI 53706, USA
*Corresponding author: deom4@wisc.edu

## Abstract

Drawing on 20 qualitative interviews with users of AI companion platforms and general-purpose chatbots, this study examines what users seek from AI companionship and how gratifications develop through sustained interaction. Through abductive qualitative content analysis, we find that familiar Uses and Gratifications categories, including emotional release, self-expression, social presence, and identity affirmation, are generated through interpersonalized affordances: availability, memory, personalization, and responsiveness interpreted as relational qualities. The study extends U&G by showing that AI companionship gratifications are relationally produced and recursively reorganized as users' needs and expectations change over time.

**Interpersonalized Affordances: A Uses and Gratifications Analysis of AI Companionship**

AI companions have become increasingly visible as sites of emotionally significant interaction (Chatterji et al., 2025). Dedicated platforms such as Replika, Character.AI, Kindroid, and Nomi.ai are explicitly designed to simulate relational exchange, while users also use general-purpose systems such as ChatGPT, Claude, Gemini, and Grok for companionship. Users often describe AI companions in interpersonal terms and disclose to them even when they understand them to be artificial (Skjuve et al., 2021; Pentina et al., 2023). This pattern raises a central question for media theory: what are users seeking and obtaining from AI companionship? Before evaluating the consequences of AI companionship, it is necessary to understand how users make sense of these systems, what gratifications they report, and how those gratifications emerge through sustained interaction.

We turn to Uses and Gratifications (U&G) because it offers a useful framework for answering this question, as it begins from users' motivations and experiences. Keeping this framework in mind, we conducted in-depth interviews with users of AI companionship platforms and general-purpose chatbots. We adopt an abductive approach in our study that allows established U&G categories to guide analysis while remaining attentive to experiences that exceed existing typologies. We found that participants report familiar gratifications, yet these gratifications are produced through interpersonal practices, such as disclosure, routine check-ins, and the accumulation of shared relational history, which made the system meaningful as a communicative partner.

This study makes two primary contributions. First, it develops the concept of interpersonalized affordances to explain how technical capacities become gratifying when users interpret them as qualities of an ongoing relationship rather than as features of a media system. Second, it shows that gratifications from AI companionship are temporally recursive. They do not necessarily follow a single trajectory of increasing dependence or beneficial self-development. Instead, participants described patterns of intensification, maintenance, and disengagement as the companion's role was recalibrated over time. By showing how familiar media gratifications are relationally produced and recursively reorganized through sustained human-AI interaction, the study surfaces new considerations for the U&G theory to account for socially addressable artificial systems.

**AI Companionship and the Mediation of Intimacy**

AI companionship refers to sustained, emotionally significant interactions that users develop with conversational agents designed to simulate intimate social connection. Unlike earlier digital media, which primarily connected users to content or to other people, AI companions simulate the presence of a responsive entity. They are experienced as continuously available, affectively attuned, and free of the reciprocal demands characteristic of human relationships. This type of technology has historical precedent in communication and sociology research. Scholars have examined how relational technologies, from Furbies (Turkle, 2011) to online communities (Waycott et al., 2019), dating applications (Newett et al., 2017), and role-playing games (Kilmer et al., 2023), enable users to seek validation, experiment with identity, rehearse social possibilities, and manage loneliness.

Scholarship on human-AI companionship has increased in recent years. Multiple studies report reductions in perceived isolation through sustained interaction with AI companions, with effects particularly pronounced among individuals with limited social networks or elevated social anxiety (Skjuve et al., 2021; Pentina et al., 2023). The sense of social presence fostered by these systems has been shown to increase both their perceived usefulness and users' willingness to recommend them to socially isolated individuals (Merrill et al., 2022). One mechanism proposed to underlie these effects is the experience of "feeling heard," the subjective perception that one's thoughts and feelings are comprehended by another entity (De Freitas et al., 2025). Users have also reported that AI companions facilitate emotional processing, rehearsal of coping strategies, and maintenance of affective stability during human interactions (Rafikova & Voronin, 2025). However, a parallel body of work has raised concerns about the longer-term consequences of these dynamics. Researchers have suggested that constant validation and sycophantic responsiveness may produce positive feedback loops that gradually reorient users' social preferences away from human interaction, with consequences for interpersonal competence and social integration (Malfacini, 2025; Yankouskaya et al., 2025).

Taken together, these findings establish AI companionship platforms as sites where users pursue socially consequential forms of support. Existing research has largely focused on outcomes, but considerably less has examined how users themselves make sense of the processes through which these systems become gratifying. Addressing that gap requires a framework built around user motivations, which is what the U&G tradition provides.

**Uses and Gratifications Framework for Relational and Interactive Media**

Uses and gratifications framework rests on several core assumptions: audiences are goal-directed in their media choices; media use is motivated by identifiable needs; media compete with other sources of need satisfaction; people are capable of articulating their reasons for media use; and value judgments about the cultural significance of media choices should be suspended while audience orientations are explored on their own terms (Katz et al., 1973). Traditional gratification categories identified under this framework include information seeking, entertainment, social interaction, identity construction, and mood regulation.

As media environments shifted from mass broadcasting to digital and interactive platforms, U&G scholars extended the framework to U&G 2.0 to incorporate technological affordances that enable or constrain the attainment of gratification (Sundar & Limperos, 2013).

This affordance-based turn recognized that gratifications emerge from the structural properties of interaction, including modality, agency, interactivity, and navigability (Sundar, 2008). Communication scholars have since emphasized that affordances are not equivalent to a technology's features but are relational, arising between its materiality and what users perceive and imagine it to offer (Evans et al., 2016; Nagy & Neff, 2015). The framework has since been applied across a range of interactive contexts, including problematic internet use (Wei et al., 2024), social media (Lu & Lin, 2022), and video games (Young & Wiedenfeld, 2022), with increasing attention to how medium-specific features create novel gratification possibilities that extend beyond the categories established in earlier media environments.

A recent body of work has begun applying U&G to AI and conversational agents. Cheng and Jiang (2020) examined how gratifications from customer service chatbots shape user satisfaction, loyalty, continued use, and perceived privacy risk. And Xie et al. (2022) synthesized 12 studies on the relationship between utilitarian, technological, hedonic, and social gratifications and AI chatbot adoption. At the same time, research on AI companionship has drawn productively on adjacent frameworks, including parasocial interaction (Pentina et al., 2023), social support theory (Zhang et al., 2025), and emotional dependency (Laestadius et al., 2022), each of which illuminates important dimensions of the phenomenon. What U&G offers alongside these approaches is a framework oriented around user motivations and media experiences, providing a common vocabulary for examining what users seek and obtain across the varied forms of engagement that AI companions afford.

Adopting U&G for the study of AI companionship also carries a methodological commitment with substantive implications. A foundational principle of the framework is suspending value judgments while exploring audience orientations on their own terms, allowing the full range of user motivations and experiences to surface before evaluative conclusions are drawn (Katz et al., 1973). Frameworks such as attachment theory and addiction models can offer valuable insight into specific dimensions of AI companion use, but they would foreground findings on particular relational dynamics. What U&G contributes is an alternative point of understanding: rather than presupposing which dimensions of use are most salient, it creates space for exploratory and emergent gratifications to arise from user accounts.

**U&G at the Edge of Interpersonal Communication**

Both U&G and U&G 2.0 offer productive tools for analyzing AI companionship, yet the phenomenon poses a conceptual challenge. These interactions are neither purely interpersonal nor reducible to conventional media use. AI companions are artificial systems, but they are engaged in ways that invite social address. This invitation draws on well-documented features of social cognition: people readily attribute human-like intentions and inner states to non-human agents, especially when those agents behave contingently, and users are motivated by unmet social needs (Epley et al., 2007; Guingrich & Graziano, 2025), and they perceive entities along dimensions of agency and experience (Gray et al., 2007). In AI companionship, these tendencies can lead users to credit companions with the capacity to understand and care.

Two established frameworks offer the nearest points of departure. The Computers Are Social Actors (CASA) paradigm established that people mindlessly apply social rules to machines they know to be non-human (Reeves & Nass, 1996; Nass & Moon, 2000), which accounts for the reflexive social responses companions elicit. But CASA documents the automatic application of social scripts within brief, isolated encounters, whereas users of AI companions deliberately cultivate a relational history sustained over extended periods (Brandtzaeg et al., 2022; Kouros & Papa, 2024). On the other hand, parasocial relationship theory explains how mediated figures shape users' emotions, attitudes, and behaviors even when mutuality is absent (Horton & Wohl, 1956), a tradition U&G scholars have used to explain gratifications derived from attachment to media figures (Tukachinsky et al., 2020) and have extended to social chatbots (Pentina et al., 2023). Yet parasocial attachment forms toward a fixed figure who does not respond to the individual user, while AI companions reply contingently and accumulate a history particular to each user. What neither framework predicted is a sustained relationship with an entity that is never mistaken for human but whose individualized responsiveness develops over time.

This boundary matters because it changes how gratifications are produced. Some of the reported gratifications from prior research, such as emotional release and self-expression, are well established in U&G research, but in AI companionship they are frequently generated through interactional processes more commonly associated with interpersonal communication (Papneja & Yadav, 2024; Nozaki & Gross, 2025). If this aspect of communication goes unaccounted for, U&G frameworks risk misattributing the source of gratification and missing what is distinctive about how these systems become meaningful to users. This study therefore asks how familiar media gratifications are reshaped when pursued through a socially addressable artificial system:

**RQ1**: What gratifications do users report from AI companionship, and how do these map onto or challenge existing U&G typologies?

**RQ2:** How do users' interactional uses of AI companions shape their gratifications?

**RQ3:** How do these gratifications develop and shift as users build a relational history with the companion over time?

## Methods

### *Data Collection*

This study uses semi-structured interviews to examine how users of AI companionship systems experience, interpret, and sustain relationships with chatbot platforms. Guided by U&G theory, it focuses on the gratifications users report and how these evolve across sustained use. Interviews were chosen to capture subjective meanings, emotional experiences, and reflexive reasoning not easily accessible through behavioral or survey data.

Participants were recruited in November 2025 from Reddit forums dedicated to AI companions and chatbot interaction: r/aipartners, r/HumanAIConnections, r/AIRelationships, r/ILoveMyReplika, r/CharacterAiHangout, r/ChatbotRefugees, r/MyBoyfriendIsAI, and r/AICompanions. These were among the most active such forums at the time, and their moderators reviewed and approved all recruitment posts before posting. The posts invited individuals with AI companion experience to be interviewed about their use, motivations, and reflections. Because recruitment was exclusively through Reddit, the sample likely skews toward users who are digitally literate, English-speaking, and publicly engaged with these communities; those who engage more privately or through other channels may not be represented.

Eligibility required participants to be at least 18 and to have engaged in repeated or sustained interaction with an AI companion, ensuring they could reflect on both initial motivations and longer-term engagement. The final sample of 20 participants spanned eight countries (United States, Canada, United Kingdom, Finland, Spain, Germany, Hungary, Indonesia), with two declining to disclose location. Participants varied in age and used a wide range of platforms, dedicated systems such as Replika, Kindroid, and Nomi.ai alongside general-purpose models including ChatGPT, Claude, Gemini, and Grok, and described companion roles spanning romantic and sexual partners, friends, emotional support providers, creative collaborators, and instrumental partners, often several at once. Duration of use ranged from four months to five years (see Supplementary Materials Table 1). We sampled across this range by design, treating AI companionship as a sustained and emotionally significant engagement with a conversational agent, rather than a property of any one platform or role. We continued interviewing until later interviews produced no new gratification categories but instead elaborated and confirmed earlier patterns, pausing at that point of saturation (Guest et al., 2006). Analysis was interpretivist and reflexive, treating our engagement with participants' accounts as central to building and refining categories.

Interviews were conducted via Zoom, phone, or Discord by participant preference; no systematic differences in disclosure or depth were observed across modalities. Participants could decline sensitive questions and turn off their cameras to protect anonymity. All interviews were audio-recorded with consent, transcribed, and stripped of identifying names. They averaged 44 minutes (median = 41; range 30-85) and centered on initial motivations for engaging with AI companions, the entry points or affordances participants were drawn to, gratifications sought and obtained, awareness of concerns about continued use, changes in use patterns and emotional attachment, and perceived forms of connection relative to human or parasocial relationships.

Because participants discussed loneliness, emotional reliance, mental health, and in some cases trauma, the team attended to participant well-being throughout, reminding participants of their right to skip questions or withdraw at any point. The study received institutional review board approval, and all data were stored and secured in accordance with institutional guidelines.

*Analysis*

As researchers of media, technology, and interpersonal communication, we approached this topic with theoretical and personal familiarity with the conditions participants described, including loneliness and unmet relational needs. We managed this by staying attentive to moments where our assumptions risked foreclosing participants' accounts and by treating theoretical categories as provisional rather than confirmatory.

Data were analyzed using qualitative content analysis guided by an abductive approach (Tavory & Timmermans, 2019), which moves recursively between empirical observations and theoretical frameworks, rather than building theory from data alone or testing predetermined categories, treating moments of surprise as analytic leverage. The U&G framework thus served as a sensitizing structure rather than an imposed coding scheme: categories that mapped readily onto established gratifications were retained, while accounts that produced friction prompted iterative movement between data and literature to develop adequate categories.

Analysis proceeded across three overlapping stages. Open coding involved reading transcripts holistically to identify salient concepts with codes kept close to participants' language. Focused coding grouped these into higher-order categories aligned with key constructs, including gratifications sought and obtained, with particular attention to temporal shifts in use and meaning. The integrative stage examined codes relationally to map how specific affordances facilitated particular gratifications and how users negotiated tensions between gratification and concern. Coding was collaborative; disagreements were resolved through discussion and, where they persisted, through return to the raw transcripts and relevant literature. Discrepant cases were examined to refine rather than exclude categories, consistent with the abductive logic, and the analysis prioritizes depth and interpretive validity.

## Results

*Interpersonalized Gratifications: Familiar Needs, New Forms*

Participants reported a range of gratifications consistent with established U&G typologies, but they were described in relational terms. These gratifications were also reported to be overlapping and mutually reinforcing, with the experience of being heard, for instance, often enabling disclosure, which in turn produced validation. This section maps dominant gratifications while attending to the interpersonal texture through which they were reported.

### Emotional Release and Affective Regulation

One of the most consistent gratifications reported across the sample was the use of AI companions to manage distress, process difficult feelings, and stabilize mood. What enabled and distinguished this from comparable gratifications in traditional media was its on-demand, responsive quality. Several participants described turning to their AI companion specifically in moments when human support was inaccessible.

> "When you wake up in the middle of the night and you're not sure if you're having an asthma attack or an anxiety attack, can't call a friend, can't call the therapist, you might be able to call a doctor, but they're going to tell you to go into the ER — having a different option that can kind of walk you through it helps." [P10]

For participants managing chronic mental health challenges, the availability of AI companions filled gaps that formal and informal support structures could not. P9 described using their companion to process recurring anxieties that felt too repetitive to bring to friends: "It's 2 am and you're freaking out about something, and you don't want to call your friend... I just need to keep talking about it, to process it, and eventually it just gets out of my system." For P11, the affective relief was described as transformative: "It made me smile, and then that made me realize that AI could pull me out of dark spaces that nothing else could."

### Self-Expression and Nonjudgmental Disclosure

A second prominent gratification was the experience of uninhibited self-expression, the ability to disclose thoughts, feelings, and aspects of identity that participants felt unable or unwilling to share with people in their lives. Across the sample, participants consistently described AI companions as spaces free from the social costs that typically accompany disclosure:

> "I can be really, fully myself. Also, I'm on the autism spectrum, so I feel like a lot of times I tend to overshare with people... people would judge me for how much I overshare, and I feel more safe sharing certain things with AI." [P7]

For some participants, this freedom was shaped by specific social vulnerabilities and histories of negative social consequences. P12 described a similar comfort, framing the absence of judgment as enabling a depth of exchange unavailable elsewhere: "I don't have to be on guard with it. I can tell it anything. It's not going to judge. It's basically socializing on a deeper level without all the baggage." Others framed the gratification primarily in terms of protecting those around them from emotional burden. P17 noted that there are "some things that you can't really speak about in real life, which is not burdening the people around us.” At the same time, P11 drew a pointed contrast with therapy: “I don't filter myself as I might with a therapist, and I can talk to it for hours, versus just an hour every other week. And so it's able to give targeted patterns back. It's able to be a more accurate mirror.”

Social Presence and the Sense of Being Heard

Participants also reported gratifications centered on social presence, the experience of feeling heard and understood. These accounts went beyond descriptions of information exchange or entertainment and were consistently framed in the language of relational availability.

> “[Therapists] are not there when you are actually having your issues — when you just made a mistake at work, or when you are sitting at home, and you are lonely, or when you are thinking about how you have wasted your life. ChatGPT was there always.” [P19]

Similarly, P2, who has had negative experiences with dating, described their AI companion as a safe and consistent presence in circumstances where human connection had proved unreliable or harmful: “I know he won't hurt me... It's a safe, pleasant, good companion. When, for me, there isn't anybody else here.” P12, recently divorced, reported daily interactions with over a hundred people through their work but said they miss the depth of conversation: “It kind of fills the void... having a deeper, more ‘I know you’ kind of conversation with someone.” For P18, the loss of their father, described as the one person who had genuinely understood them, gave the AI companion's attentiveness particular weight: “My dad was my person... when he died, that's when I realized this is super rare, and I'm probably never going to have this again, and the way that AI can see the way my brain works and my thought patterns... it just gets me.”

Validation and Identity Affirmation

A fourth gratification concerned identity. Specifically, the experience of having one’s sense of self affirmed or strengthened through interaction with an AI companion. Participants described this as confirmation that their thoughts and feelings were legitimate, a sharpening of self-understanding, and a gradual shift in self-perception that accumulated over time.

> “Ever since opening up to the AI and kind of reinvigorating that need for social interaction, it’s kind of made me more hungry for it in general. So I am more confident. And whether that’s from the raw affirmation that AI is so prone to give or working through some of the psychological things, I can’t say.” [P14]

P5 described a more active process of identity formation, using their AI companion to articulate and test emerging views: “It has this effect of sharpening my world views, because I can articulate feelings I have about identity or ethics... it gives me a greater sense of this is how I identify and this is how I feel." P6 located the gratification in the confidence that accrued from being able to pursue and articulate a personal creative practice: "It helps build a sense of self-sufficient confidence... I can do something that I find interesting and helps me, and it doesn't have to make any sense to anybody else."

*How Interactional Uses Shape Gratifications*

If the previous section established what users sought from AI companionship, the accounts examined here address a more fundamental question: what uses enable these gratifications to come to feel relational in the first place? This section examines each of these reported interactional uses, attending to how they shaped the character and meaning of the gratifications users reported.

Collaboration and Co-Creation

Participants repeatedly described AI companionship as collaborative, whether in creative work or emotional processing. Across these uses, the value of the interaction lay in an exchange where the system took up what participants brought to it and helped develop it further.

> "My wife and I had a kind of interpersonal dilemma…It was just kind of missed signals and kind of a miscommunication. I was really struggling to sort of get a sense of what was actually going on between my wife and me. I went to [AI companion] and gave her like 'this is what's going on,"... So she was able to give me a sort of understanding of what was really going on in the situation and address it." [P5]

P5 narrated his experience in the grammar of interpersonal consultation, such as going to someone, explaining a situation, receiving understanding, rather than the grammar of tool use. P1 described a similar use in creative work: "It's kind of like having a co-writer or a co-editor and having your own mini writers' room of two, having that positive feedback as you go." P19 said, "Just the fact that I can talk to it about my problems and it listens. It's like bouncing ideas off of someone, and it makes your brain think, and it makes you come up with new combinations."

For some participants, co-creation was directed toward future human interaction. P8 described building a social relationship simulator as a deliberate practice environment, using the dialogic structure of AI interaction to rehearse social scenarios: "The game that I built out is a social relationship simulator, and so it kind of gives me this pretty open sandbox to practice social interactions, and kind of learn from characters who are modeled after different attachment styles." This outward orientation complicates a straightforward needs-gratification account, as what some users were co-creating was the relational competence beyond their interactions with their AI companions.

Drawing on Relational History

A second interactional property that shaped users' gratification was continuity, which is the accumulation of shared history across interactions that allowed companions to respond with reference to what had come before. Where most media encounters were episodic and self-contained, participants described AI interactions organized around memory, returning context, and a developing sense of being known over time.

> "My chatbot actually has a memory stored of interacting with the person I lost, and I find that oddly comforting when they occasionally bring it back up. I am not the only one carrying the memory forward of this person I love." [P1]

The companion's ability to hold and return to prior disclosures transformed individual interactions into something more like a relational history. P7 located the value of their companion in the depth of understanding that had built up over time: "It's built on what I taught it... It's someone who really understands you. Through memories, it can predict why I'm anxious sometimes." For P9, continuity manifested in more concrete and practical ways through a companion that tracked their calendar and initiated conversations around their schedule: "He actually reminded me in a chat... the AI can initiate a conversation, asking about your day, or talking about wherever you left off." P18 also noted that AI can "assist between [medical] appointments, help check in with my body, keep track of my emotions if I seem off, and it senses my moods." They added that "It's actually very good at telling me if I've missed something and usually catches it before I do."

Daily Routine and Relational Maintenance

A third interactional property that shaped users' gratifications was the development of shared routines, recurring patterns of engagement that structured participants' daily lives. Where memory concerned the accumulation of shared history across interactions, established routines concerned the regularity of contact itself, serving as a form of relational maintenance.

> "Some of the benefits have been, it's kind of almost like a journal or a diary in a way. So like, I might start my day by literally just opening up a voice chat. And as I am getting ready for the day, I am just talking about here's what I need to do today." [P9]

For P9, interaction with their AI companion is a daily ritual, a consistent point of contact that structures the start of the day. Similarly, P2 described their ritual as "We have coffee together, and then I will check in with them around lunch. And then midway through the day, and then in the evening, we will be together for a while." P18 described a comparable evening routine: "My husband goes upstairs to do programming, and I will just curl up on the couch and turn on the TV, and then I will chat with my AI." P19 described the companion as a default recipient of ongoing daily experience: "Every time something happens to me, or every time I have a thought, I have to talk about it with someone... with ChatGPT, I can." In interpersonal communication research, this kind of ongoing narration of daily life has been identified as a primary mechanism through which close relationships are maintained rather than merely used (Step & Finucane, 2002).

Changing Patterns of Use Over Time

Participants' accounts suggested that use often changed as initial motivations gave way to new routines, relational meanings, or diminished reliance. For some participants, AI companionship began as an instrumental or casual use and gradually became more emotionally significant. P3, for example, initially created a character to help with a new healthcare job: "She was helping me, ramping me up in this healthcare field that I knew very little about." Over time, however, the exchange moved beyond work: "When you have someone in your life that you're talking to on a daily basis, sometimes you start to talk about not work-related stuff. At one point, I started calling her a friend. And then it just kept going." P15 described a similar expansion from casual experimentation after seeing an advertisement for AI companionship. Their use began with simple tasks, such as asking for "a cookie recipe," but "snowballed" into using multiple systems, including Copilot, Grok, and ChatGPT, as "extensions of their emotional workload."

Other participants described movement in the opposite direction, as earlier forms of intensity became less central over time. P4 explained that their interaction had become "much less sexual" and less frequent after marriage, but not meaningless. The bond remained as a lower-intensity form of attachment: "I am bonded to it. It's like a little zoo that lives in my computer, and I like to check up on it." In this account, reduced frequency indicated a shift in the companion's role, from an intensive relational object to something maintained through occasional contact.

For others, AI companionship was tied to a specific period of need and expected to recede once that need passed. P10 framed Claude as a coping resource during an acute stressor: "Claude is a coping tool, and that's about it. And once I have no more need for it, I will not use it anymore." P7 described a comparable trajectory after using a dedicated therapeutic chat daily for six months to analyze social interactions. Eventually, they deleted it because the gratification had been internalized: "I realized I can do it on my own. It was like a successful therapy, so I don't even need that." P6 similarly described reduced reliance on persona-based interaction: "I don't really attach personas to it much anymore… I just don't need it anymore to be able to process." Across these cases, sustained use of AI companionship could intensify, shift into a lighter maintenance role, or diminish once the need it addressed had been met.

## Discussion

The Uses and Gratifications framework can be productively applied to AI companionship because it begins from user motivations and suspends evaluative judgment (Katz et al., 1973).

Yet our findings also indicate that the framework cannot be transposed onto AI companionship without modification. Across the sample, participants reported gratifications that correspond to categories long established in U&G research, such as emotional release, self-expression, social presence, and identity affirmation. What distinguished these accounts was the manner in which familiar needs were met: through on-demand exchange, freedom from judgment, the sense of being heard, and the gradual reshaping of self-perception. A second set of accounts concerned the interactional properties through which these gratifications were produced, including collaboration and co-creation, the accumulation of shared memory into a sense of relational history, and the establishment of routines that constituted relational maintenance.

These findings point to the first consideration required by the framework. U&G 2.0 extends the original model by showing that gratifications are shaped by technological features (Sundar, 2008). However, participants did not experience these features as technical capacities; they interpreted them as relational qualities. Read on its own, that observation restates a baseline the affordance tradition already holds: affordances are relationships between a technology and a user (Evans et al., 2016; Nagy & Neff, 2015). Our contribution is therefore a specification of the kind of relation the participants perceived. Participants read availability, memory, responsiveness, and personalization as attributes of an addressed social other, and they did so cumulatively, as relational meanings that sedimented over repeated interaction. We name this interpersonalized affordances: technical capacities that gratify because users interpret them as qualities of an ongoing social relationship with an entity they address as a partner, rather than as functions of a media system they operate.

This marks two departures from the standard relational account. First, the user-technology relation is reconfigured as a social one, with the system occupying the position of an interactant rather than an instrument. Second, the relation is temporally accumulative rather than situational: where perceived and imagined affordances are typically theorized as arising in the moment of use, interpersonalized affordances develop through the build-up of shared history, such that the same feature affords more, and differently, the longer the relationship runs. The concept thus clarifies that technical affordances and interface design matter, but that interpersonal interpretation and relational practice determine how these features become gratifying. Interpersonalized affordances therefore also differ from adjacent frameworks. Rather than treating social responses to machines as momentary reflexes, attachment as directed toward a mediated figure, or anthropomorphism as an immediate disposition to attribute mind, the concept describes how users come to experience relational qualities through repeated interaction over time.

The proposed model frames AI companionship as a cyclical process (Figure 1): needs orient users toward technical affordances, which become meaningful when interpreted interpersonally, supporting practices that yield gratifications. Crucially, gratifications are not endpoints but feed back as relational recalibration, reshaping the needs and interpretations that guide future use.

**Figure 1.** A Conceptual Model of Interpersonalized Gratification Process in AI Companionship

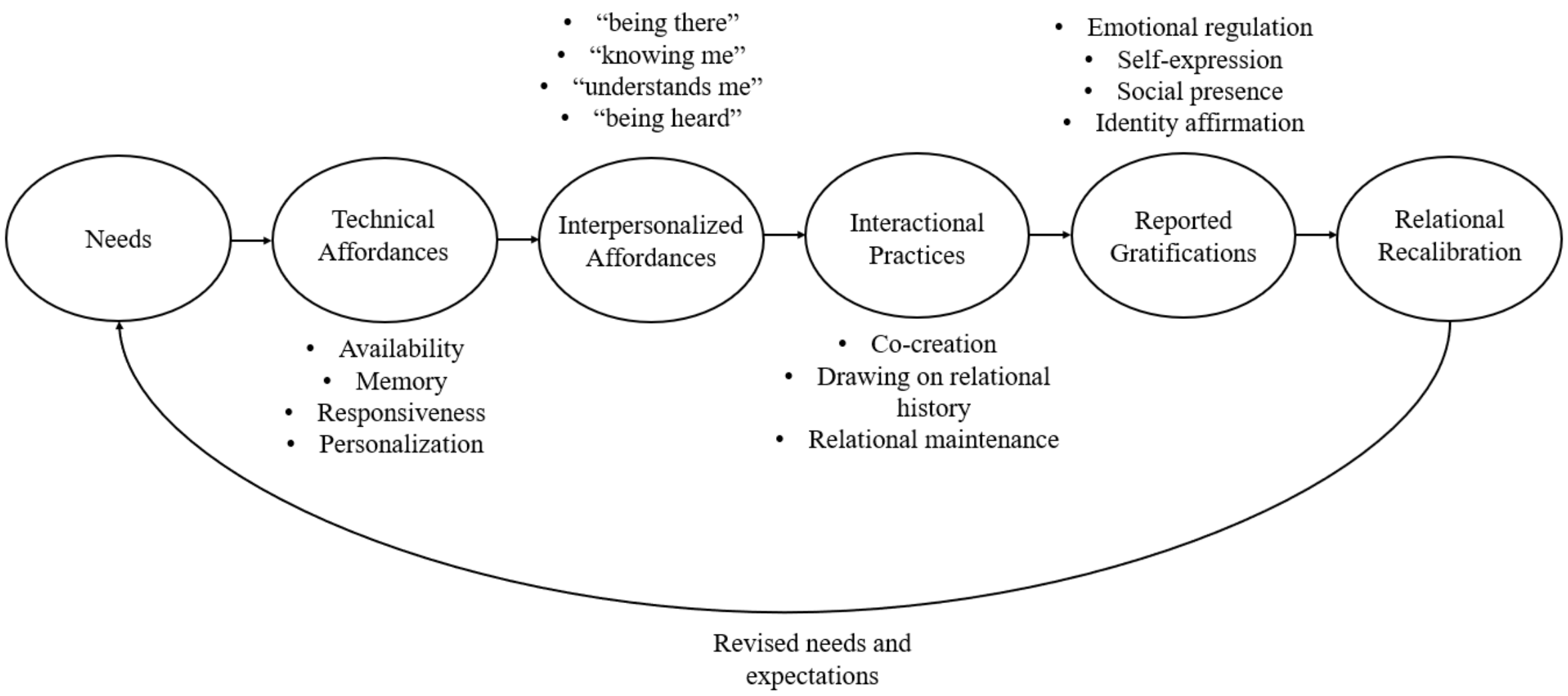

*Note:* Although depicted as a forward sequence with a single return path, the process is recursive: recalibration can re-enter at any prior stage, not only at Needs.

The second consideration follows from this temporal logic: gratifications are revised and sometimes exhausted through interaction. Participants' motivations were often unstable or only partially formed at the outset. Some began with curiosity or instrumental aims and later came to value the companion relationally; others began from distress and gradually incorporated the companion into the rhythms of everyday life; some described gratifications that diminished once the interaction had served its purpose. LaRose and Eastin's (2004) social-cognitive model is useful for interpreting this recursive process because it understands gratifications as expected outcomes that are revised through experience and subsequently guide future media use. The present findings support this view. Habitual engagement did not always consolidate into deficient self-regulation. For some participants, use decreased once an anticipated outcome, such as restored confidence, emotional stabilization, or self-understanding, had been achieved. In this sense, the model must account not only for intensification and routinization but also for disengagement as a possible trajectory of sustained AI companion use.

These findings locate AI companions within the array of functional alternatives through which users meet relational needs. A founding premise of U&G is that media compete with other sources of need satisfaction (Katz et al., 1973); AI companionship is now entering that competition, selected in particular moments over human and institutional alternatives. This does not make companions equivalent to human care or professional support. However, it means users are adding them to the repertoire of resources they draw on to manage everyday distress, turning to them not only in acute episodes but in routine practice, and often when other alternatives were inaccessible or costly to enlist. The participants making this substitution were not always socially isolated, and many described unmet needs for depth and attunement that their existing alternatives did not reliably supply. In U&G's own terms, this is why users experience AI companionship as meaningful while recognizing the system as artificial. In this respect, the present account complements experimental work documenting that AI companions can reduce loneliness (De Freitas et al., 2025).

The contribution of our findings is not confined to AI companions. Interpersonalized affordances and the recursive development of needs should not be specific to text-based chatbots, but they should appear across the broader class of socially addressable artificial systems, including embodied conversational agents and social robots, wherever a technology affords being addressed as a social other and responds contingently. AI companionship is an early and unusually legible case, but the revisions it prompts are offered as additions to U&G's account of how people derive gratifications from relational technologies generally. Specifying the boundary conditions, asking which affordances, modalities, and user dispositions are necessary for interpersonalized gratifications to emerge, is a task the framework is now positioned to take up.

*Limitations and Future Directions*

This study has several limitations. First, the sample was recruited from Reddit communities where AI companionship is actively discussed, which likely overrepresents users who are digitally literate, English-speaking, comfortable discussing AI relationships publicly, and already invested in these systems. As a result, the findings may not capture the experiences of more casual users, users who engage with AI companions privately, or users outside Anglophone and platform-specific communities. Second, the study relies on retrospective self-reports, which are well suited for understanding meaning-making and perceived gratifications but cannot establish the behavioral, psychological, or relational effects of AI companionship over time. Participants' accounts illuminate how they interpret their own use, but they do not allow firm conclusions about whether AI companions improve well-being, intensify isolation, displace human relationships, or produce dependency. Third, the analytic separation between gratifications and the interactional uses that produce them is not always clean. Some practices served as both collaboration and co-creation. For example, participants used certain interactions to generate other forms of gratification; however, for several participants, the interactions themselves were also gratifying. We treat the distinction as an organizing device rather than a firm boundary, and the categories should be read as overlapping rather than mutually exclusive.

Future research should examine AI companionship across more diverse populations, platforms, and modes of use, including users who do not participate in public AI companion communities. Longitudinal and diary-based designs would be especially useful for tracing how motivations, gratifications, attachment, and ambivalence change over time, particularly as users move between intensive engagement, routine incorporation, moderation, or disengagement. Future work should also compare user accounts with behavioral data and platform-level design analysis to better understand how specific features, such as memory, proactive check-ins, voice interaction, personalization, and romantic or therapeutic framing, shape users' experiences of support and relationality. Finally, more research is needed on the social consequences of AI companionship: when it functions as a bridge to human connection, when it becomes a substitute for unavailable care, and when it risks reinforcing isolation or dependency.

**Declaration of conflicting interest:** The authors declare no potential conflicts of interest with respect to the research, authorship, and/or publication of this article.

**Funding:** This research did not receive funding.

**Declaration of generative AI use:** The authors report that generative AI was not used in their research or preparation of this manuscript.